\documentclass[aps,preprint]{revtex4}%
\usepackage{amsfonts}
\usepackage{amsmath}
\usepackage{amssymb}
\usepackage{graphicx}%
\begin{document}
\preprint{ }
\title[ ]{History-dependent relaxation and the energy scale of correlation in the Electron-Glass}
\author{Z. Ovadyahu}
\affiliation{The Hebrew University, Jerusalem 91904, Israel }
\author{$.$M. Pollak}
\affiliation{Department of Physics, University of California, Riverside CA 92651, USA}

\begin{abstract}
We present an experimental study of the energy-relaxation in
Anderson-insulating indium-oxide films excited far from equilibrium. In
particular, we focus on the effects of history on the relaxation of the excess
conductance $\Delta$G. The natural relaxation law of $\Delta$G is logarithmic,
namely $\Delta$G$\varpropto$log(t). This may be observed over more than five
decades following, for example, cool-quenching the sample from high
temperatures. On the other hand, when the system is excited from a state
S$_{o}$ in which it has not fully reached equilibrium to a state S$_{n}$, the
ensuing relaxation law is logarithmic only over time t shorter than the time
t$_{w}$ it spent in S$_{o}$. For times t$\geq$t$_{w}$ $\Delta$G(t) show
systematic deviation from the logarithmic dependence. It was previously shown
that when the energy imparted to the system in the excitation process is
small, this leads to $\Delta$G$\varpropto$P(t/t$_{w}$) (simple-aging). Here we
test the conjecture that `simple-aging' is related to a symmetry in the
relaxation dynamics in S$_{o}$ and S$_{n}$. This is done by using a new
experimental procedure that is more sensitive to deviations in the relaxation
dynamics. It is shown that simple-aging may still be obeyed (albeit with a
modified P(t/t$_{w}$)) even when the symmetry of relaxation in S$_{o}$ and
S$_{n}$ is perturbed by a certain degree. The implications of these findings
to the question of aging, and the energy scale associated with correlations
are discussed.


\end{abstract}
\maketitle

\section{ Introduction}

Recently there were several reports on the non-ergodic transport properties of
Anderson insulators.\cite{1,2,3,4} These were interpreted as evidence for a
glassy phase that was theoretically predicted by several authors.\cite{5} When
excited from equilibrium, the conductance of such systems increases - a
property inherent to the hopping system.\cite{1} The excess conductance
$\Delta$G persists for long times (in some cases, days) after the excitation.
Vaknin et al \cite{6} argued that such extended relaxation times as well as
other glassy aspects are associated with the presence of strong inter-carrier
interactions.\cite{7} The various memory effects exhibited by these systems
are difficult to explain unless electron-electron correlations play a role.

An intriguing memory effect, common to many other glasses, is aging. \cite{8}
This term usually refers to the response of a system when it is under the
influence of two differently imposed conditions, the first lasting a
macroscopic time t$_{w}$, the next (immediately following) a time t. The
system is said to age if the response depends on both t$_{w}$ and t. That is
in \textit{fundamental} contrast to ergodic systems where the response depends
only on t. In the electron glass this may be observed in $\Delta$G(t) by,
e.g., suddenly changing the gate voltage (employing a MOSFET structure). It
turns out that the aging function $\Delta$G(t,t$_{w}$) in this case is just a
function of t/t$_{w}$.\cite{9}. This so called `simple' or `full' aging
behavior has been rarely observed in such a clean form in any other glassy
system. It was conjectured that simple aging is associated with the identical
dynamics in the `old' and `new' states involved in the experiment. Indeed, the
`anti-symmetric' behavior of $\Delta$G(t) in the `old' and `new' states, and
their logarithmic form are sufficient to guarantee simple-aging for t%
$<$%
t$_{w}$.\cite{10} In this paper, we test this conjecture by exploring the
changes in the aging function caused by situations where the above symmetry
begins to break down. This symmetry is probed by a new experimental procedure
described in section III. It is shown that the dynamics in the `old' state is
affected by both, the time it takes to switch the external condition that
imposes the `new' state, and by the energy difference between the two states.
Both contribute to asymmetry in the dynamics. We compare these changes with
the respective modifications in the aging function and discuss the
implications as to the question of the energy scale for the correlations that
presumably control the glassy behavior in the system.

\section{Experimental}

Samples used in this study are thin films (50\AA \ thick) of crystalline
indium-oxide e-gun evaporated on 140$\mu$m cover glass. Gold film (500\AA
\ thick) was evaporated on the backside of the glass and served as the gate
electrode. Conductivity of the samples was measured using a two terminal ac
technique employing a 1211-ITHACO current preamplifier and a PAR-124A lock-in
amplifier. All measurements reported here were performed with the samples
immersed in liquid helium at T=4.11K held by a 100 liters storage-dewar. The
ac voltage bias was small enough to ensure Ohmic conditions. Fuller details of
sample preparation, characterization, and measurements techniques are given
elsewhere.\cite{11} Four different batches of samples were used in the present
study. Sample size was typically 1x1 mm and its sheet resistance R$_{\square}%
$was 30-80 M$\Omega$ at 4K.

\section{Results and discussion}

Figure 1 shows the dependence of the sample conductance G following a quench
from a high temperature to the measurement temperature T$_{m}$. The observed
G(t) is seen to obey a logarithmic law over more than five decades. Figure 2
illustrates a similar G(t) for the same sample measured after V$_{g}$ was
changed from -50V to +50V. We shall show that the logarithmic dependence
characterizes the approach to equilibrium of the electron glass when no
history intervenes. The notion of history will become clear below. For now it
is emphasized that in both cases depicted in figures 1 and 2, the relaxation
is monitored under conditions where the system exhibits no signature of the
past; In the first case the system is relaxing from a high-energy state
(presumably ergodic, e.g., above its glass temperature), and thus have no long
term memory of its old state. In the second case, as will be demonstrated, the
system does have a memory of the old state but the signature of this memory
does not appear in G(t) during the time of the measurement. This is so because
the time the system spent in the old state was much longer than the time over
which G(t) was monitored during the experiment shown in figure 2. Indeed, when
the time over which G(t) is recorded is comparable or longer than the
\textquotedblleft waiting\textquotedblright\ time t$_{w}$ the system was
allowed to equilibrate at the old state, a clear deviation from logarithmic
dependence can be observed. An example is shown in figure 3.

Note first that the $\Delta$G$\varpropto-$log(t) reported here extends the
observation of this relaxation law, previously reported, to include more than
five decades in time. It seems now plausible to conclude that this is the
natural (`history-free') relaxation law of the electron glass. Such a law has
been explained as being inherent to the hopping system due to its extremely
wide distribution of transition rates $\omega$.\cite{12}

The log(t) relaxation law cannot persist for arbitrarily long times even when
history does not play a role (e.g., when t$_{w}$=$\infty$) because the slope
of $\Delta$G[log(t)] must vanish asymptotically (or else G(t) will fall below
the equilibrium conductance G(0). Physically, the deviation from a log(t)
behavior is expected when t$^{-1}$ approaches the slowest rate $\omega_{min}$
in the distribution.

The relaxation in figure 3 exhibits a deviation from the logarithmic
dependence at time t$_{w}$, which is evidently much smaller than the `natural'
$\omega_{min}^{-1}$ of the system. It would thus appear that the effect of
\textquotedblleft history\textquotedblright\ on the relaxation law is to
modify the effective relaxation rate-distribution. Note however that the
observed deviation in figure 3 (that falls above the log line) is
qualitatively different from departures from log due to a rapid cut-off in a
distribution at $\omega_{min}$.

It was shown in \cite{10,13} that logarithmic behavior results from the
exponential dependence of the transition rates $\omega$ on a random variable x
with a smooth distribution. Departures from a flat distribution of x introduce
only logarithmic corrections to the 1/$\omega$ distribution of $\omega$.
However, if the distribution of x is not smooth, the deviations from log(t)
relaxation can be strong. To explain the observed behavior with a change of
the distribution mandates a rise in the distribution around t$_{w}^{-1}$ (A
rise is required for a departure \textit{above} the log line, as observed in
figure 3).

The signature of a previous excitation/relaxation process appearing in a
subsequent experiment is a peculiar property of glasses. This history
dependent relaxation is one of the aspects of `aging'. A typical aging
experiment involves the following procedure. The system is allowed to
equilibrate under a set of external conditions \{x$_{i}^{o}$\} that control
some macroscopic response function P. Then, \{x$_{i}^{o}$\} is changed to
\{x$_{i}^{n}$\} for a waiting time tw during which P(\{x$_{i}^{o}$\}) evolves
towards P(\{x$_{i}^{n}$\}). Finally, the `old' conditions \{x$_{i}^{o}$\} are
restored, and the response P is monitored versus time t starting from the
moment \{x$_{i}^{o}$\} are re-established. In a glassy system it is commonly
found that P(t) reflects t$_{w}$ (as well as \{x$_{i}^{o}$\},\{x$_{i}^{n}$\}).
Namely, P(t) is in general a function of both t and t$_{w}$. A special case of
aging is when P(t,t$_{w}$)=P(t/t$_{w}$). Such a behavior of the relaxation (so
called `simple-aging'), has been recently observed in Anderson-localized
indium-oxide thin films.\cite{9} In these experiments \{x$_{i}$\} was the
carrier concentration n (controlled via a gate voltage V$_{g}$), and the
response P was the electronic conductance G.

Vaknin et al \cite{10,13} showed that the symmetry of the two-dip experiment
(TDE) is sufficient to cause \textquotedblleft simple-aging\textquotedblright.
In the TDE, the sample is cooled to the measuring temperature with a voltage
V$_{g}^{o}$ held at the gate, and is allowed to equilibrate for several hours.
Then, a G(V$_{g}$) trace is taken by sweeping V$_{g}$ across a voltage range
straddling V$_{g}^{o}$. The resulting G(V$_{g}$) exhibits a local minimum
centered at V$_{g}^{o}$. At the end of this sweep, a new gate voltage,
V$_{g}^{n}$ (differing from V$_{g}^{o}$ by typically few tens of volts for a
100$\mu$m spacer), is applied and maintained at the gate between subsequent
V$_{g}$ sweeps that are taken consecutively at latter times (measured from the
moment V$_{g}^{n}$ was first applied). Each such sweep reveals two minima
(c.f., figure 1 in reference 9 as an example); one around V$_{g}^{o}$ (with
magnitude A1) that fades away with time, and the other at V$_{g}^{n}$ with
magnitude A2 increasing with time. It turns out that when $\delta$V$_{g}$=%
$\vert$%
V$_{g}^{n}-$V$_{g}^{o}$%
$\vert$
is not too large, A1(t) and A2(t) are `symmetric' in the sense that
A2(t)=a$\cdot$log(t) while A1(t)=A1(t=0) a$\cdot$log(t), namely, G at
V$_{g}^{o}$ evolves with time at the same rate (and opposite sign) as the
change in G at V$_{g}^{n}$.

It is easy to see that these time dependences of A1(t), A2(t) are sufficient
to guarantee `simple-aging' in the corresponding aging experiment.
Specifically, the excess G observed after V$_{g}^{o}$ is restored (at t=0)
relaxes according to the law:

$\Delta$G(t)=$\Delta$G(t=0)-a$\cdot$log($\omega$t)=a$\cdot$log($\omega$t$_{w}%
$)-a$\cdot$log ($\omega$t)= a$\cdot$log(t/t$_{w}$). Note that there are two
ingredients in this \textquotedblleft derivation\textquotedblright. The first
is the basic logarithmic relaxation (which, as discussed above, is obeyed over
a limited range in t, and thus poses a similar restriction on the relaxation
law in the aging experiment). The second is the symmetry of the A's. In the
following, we wish to look more closely into the relation between the symmetry
and simple aging. We shall describe how the symmetry is modified by the
experimental procedure, and what implications this has on the question of
simple aging.

What does the symmetry (or anti-symmetry) mean? The magnitude A of the cusp
formed in G(V$_{g}$) at a newly imposed gate voltage reflects the dynamics of
the approach towards the `new' equilibrium value of the conductance
G(V$_{g}^{n}$). Initially (t=0), G exceeds G(V$_{g}^{n}$) by a certain amount
and it slowly decays towards it. We shall refer to this process as
\textquotedblleft learning\textquotedblright. This is used here mainly to
facilitate the discussion below but note that \textquotedblleft
learning\textquotedblright\ is an appropriate name for a process where a
system adjusts to newly imposed conditions. During this learning-time, the
system begins to \textquotedblleft forget\textquotedblright\ the `old'
conditions (i.e., the cusp at V$_{g}^{o}$ fading away). The latter process is
monitored in the TDE experiment by peeking briefly from time to time at what
happens at V$_{g}^{o}$ while most of the time the system is allowed to
experience V$_{g}^{n}$. Both processes are driven by the need to adjust to
(the same) change in the external conditions, and on a microscopic level, both
involve transitions between states localized in (the same) space. The
transition rates are presumably controlled by disorder, and restrictions due
to interactions. The processes differ in that they occur in different places
on the energy scale (i.e., the energy associated with V$_{g}^{n}$ and
V$_{g}^{o}$). When this difference is small, the two processes should then be
symmetric. In the other limit, that of large energy difference, the transition
rates could be quite different thus leading to lack of symmetry in the
`learning' versus `forgetting' processes. For example, the localization length
and/or the relevant density of states may be larger at the higher energy, and
this may lead to faster dynamics.

The question is how small should $\delta$V$_{g}$ be to guarantee this
symmetry. To answer that experimentally we performed a series of measurements
using the following procedure. After allowing the sample to equilibrate under
V$_{g}^{o}$, the gate voltage was swept to another value V$_{g}^{n}$ and kept
there for a dwell-time t$_{d}$. Then, V$_{g}$ was swept back to V$_{g}^{o}$
and kept there for the same t$_{d}$, and so on. The sample conductance G was
continuously monitored during this process. Typical results of such an
experiment (which is essentially a repeated aging experiment) are shown in
figure 4. Focusing on the G at the end of each dwell-time one identifies two
sequences: G(t$_{i}$) for $\Delta$G V$_{g}$=V$_{g}^{o}$, and G(t$_{i}$) for
V$_{g}$=V$_{g}^{n}$ that increases (decreases) with t respectively. Figure 5
shows a similar G(t) for a specific $\delta$V$_{g}$ where the time is plotted
on a logarithmic scale starting from the time (plus one second, which is the
sampling time) V$_{g}^{n}$ was first established. The figure includes three
straight lines. The first is the initial relaxation curve of the conductance,
which is naturally logarithmic just as in figure 2 above. The other two are
lines connecting the G(t$_{i}$) points of the `learning' and `forgetting'
series that also appear to obey rather well logarithmic time dependence. This
is not a trivial result; each series `accumulates history' as time goes by but
the main effect of it seems to be just a modification of the logarithmic
slope. Namely, the absolute value of $\partial$G(t$_{i}$)/$\partial
$ln(t)=q$_{L,F}$ for the `learning' (q$_{L}$) and `forgetting' (q$_{F}$)
series is \textit{smaller} than the respective value for the `natural'
relaxation (q$_{n}$). Either the ratio or difference between q$_{L}$ and
q$_{F}$ may be used a measure of `learning-forgetting asymmetry'. In the
experiments described here a fixed t$_{d}$=20 seconds was used and the
asymmetry was measured as function of $\delta$V$_{g}$ and the gate-voltage
sweep-rate dV$_{g}$/dt. Typical results for q$_{L,F}$ and q$_{F}$/q$_{L}$ are
shown in figures 6 and 7 as function of $\delta$V$_{g}$ for two of the studied
samples. Looking first in figure 6 one observes that both q's increase with
$\delta$V$_{g}$, which is because the amplitude of the excess conductance
grows monotonically with $\delta$Vg. The interesting point here is that
q$_{F}$ increases faster than q$_{L}$ a trend that becomes clear above a
certain $\delta$V$_{g}$. As mentioned above, a difference in energies could
create a difference in the rate of relaxation, which in turn lead to
asymmetry. Note however that the trend observed in figure 6 is
\textit{independent of the sign} of V$_{g}^{n}-$V$_{g}^{o}$. Moreover, the
relaxation that resulted by switching V$_{g}$ from, say, -100V to +100V (after
allowing equilibration at -100V) showed the same logarithmic slope as when the
initial V$_{g}$ was +100V and relaxation was monitored after V$_{g}$ was swept
to -100V. The reason for the asymmetry at higher $\delta$V$_{g}$ is therefore
\textit{not} due to the difference in the \textit{equilibrium} energies
associated with the different V$_{g}$. We now show that the asymmetry in these
`learning-forgetting' experiments results from the combination of two factors.
The first is the finite time it takes V$_{g}$ to move between the `old' and
`new' states. During this travel time, the gate voltage is neither at
V$_{g}^{o}$ nor at V$_{g}^{n}$. This naturally contributes to the `forgetting'
process and diminishes the rate of `learning'. Namely, the conductance at both
V$_{g}^{o}$ and V$_{g}^{n}$ changes in the direction to make q$_{F}$/q$_{L}$%
$>$%
1. This can be seen in figure 7 as an explicit dependence of $\gamma$=q$_{F}%
$/q$_{L}$ on sweep rate for four fixed values of $\delta$V$_{g}$. The second
factor is less trivial. As implied by the data in figure 7, asymmetry
increases with $\delta$V$_{g}$ even when the effect due to travel-time-delay
is taken into account, i.e., for a constant t$_{d}$/[travel time]. This
contribution to the asymmetry is therefore associated with the change in the
state of the system when the gate voltage is \textit{swept through} $\delta
$V$_{g}$. Empirically, the effect of $\delta$V$_{g}$ in these experiments is
to destroy the correlations that are built during the respective t$_{d}$ for
both, the `old' or `new' state. It therefore speeds-up `forgetting' and
slows-down `learning'.

The change of V$_{g}$ affects the electrons in two different ways. One is to
change the electron population, the other is to change the random potentials
of the sites (by ez$_{i}\delta$F where z$_{i}$ is the coordinate across the
film of site i and $\delta$F is the change in the electric field induced by
$\delta$V$_{g}$). Changing the random potentials constitutes an excitation -
say the system was in the ground state prior to the change of V$_{g}$, this
state becomes an excited state just after the change because a change in the
Hamiltonian changes the ground state. $\delta$V$_{g}$ thus causes a rise in
the conductivity because quasi-particles are not fully formed in an excited
state; the motion of particles thus becomes less constrained by correlation.
We argue below that this is the dominant effect at small $\delta$V$_{g}$ and
the one that preserves the observed memory effects. The way in which that
comes about has been discussed more fully in \cite{13}

Insertion of electrons (or holes) is a relatively minor effect at small
$\delta$V$_{g}$ where excitation is smaller than the typical interaction
energy (i.e. the typical energy of formation of quasi-particles). This is so
because the energy of the particle is larger than the energy of the
quasi-particle and therefore the particle density of states at small energies
is suppressed. (For example, the Coulomb gap \cite{14} can be viewed as the
difference between the energy of particles and quasi-particles.) Furthermore,
to leave memory intact, the inserted particles must \textit{rapidly} enter the
system in just the same number and the same degree of dressing as the
particles that left the system at the lower V$_{g}$. There is clearly no
reason for this to happen.

When $\delta$V$_{g}$ causes an excitation in excess of the typical
quasi-particle energy, there is no impediment to electrons entering the system
and, since the observed onset of asymmetry happens where $\delta$V$_{g}$
corresponds to the interaction energy, we believe that this is a likely cause
for loss of memory.

We turn now to the modifications to the aging behavior as function of $\delta
$V$_{g}$ in these samples. Figure 9 shows the aging functions for two values
of $\delta$V$_{g}$ measured for the same sample using an identical set of
waiting times. At first sight, both sets of data show good scaling with
t/t$_{w}$ despite the fact that the data in 9b is evidently in the asymmetric
regime (c.f., figure 7). Therefore, `simple-aging' may apparently be observed
even when the symmetry between the `old' and `new' state is no longer obeyed.
Closer inspection reveals some differences between the small and large
$\delta$V$_{g}$ cases; The collapse of data is somewhat worse for the latter
case, and more importantly, the extrapolated value of the initial logarithmic
slope of $\Delta$G(t) crosses $\Delta$G=0 at t/t$_{w}$%
$>$%
1 as opposed to t/t$_{w}$=1 for the data in figure 9a. These differences,
however small are significant; they were consistently observed in all four
batches studied as well as in another type of experiments where a large change
in the source-drain field $\Delta$F was used to study aging \cite{15} (instead
of \textquotedblleft gating\textquotedblright\ the sample by $\delta$V$_{g}$).
This allowed a much bigger range of measurements over which asymmetry was
apparent and the extrapolation to $\Delta$G=0 kept moving up with $\Delta$F
beyond t/t$_{w}$=1 reaching a value of $\approx$7 before a pronounced lack of
a collapse of the curves was noticed.

It is illuminating to compare the relatively minor changes in the scaling of
the sample in figure 9 at large $\delta$V$_{g}$ with the corresponding
behavior of the sample studied by Vaknin et al [10] (c.f., their figure 12).
In the latter case, near perfect scaling of $\Delta$G(t/t$_{w}$) was observed
for $\delta$V$_{g}$=40V but for $\delta$V$_{g}$=240V the $\Delta$G(t/t$_{w}$)
curves completely failed to collapse. Note that in terms of $\delta$V$_{g}%
$/$\Gamma$ the sample in reference 11 (figure 12b) is $\approx$6 which is
quite similar to $\delta$V$_{g}$/$\Gamma\approx$5 for the sample in figure 9b
which still shows reasonable $\Delta$G(t/t$_{w}$) scaling. There must
apparently be a reason for the higher sensitivity of the sample studied by
Vaknin et al. Namely, the degree to which the aging behavior is modified when
$\delta$V$_{g}$/$\Gamma$%
$>$%
1 apparently involves another parameter. We suggest that the different
sensitivity in the aging behavior between the two samples may be due to the
difference in their carrier concentration n. The importance of this system
parameter in gating experiments can be understood as follows. As noted above,
the application of a `new' gate voltage has a number of consequences; the
field associated with V$_{g}$ re-shuffles the random potential of the
electronic sites. In the process, the system is re-excited (sometimes referred
to as `rejuvenation') and some loss of memory unavoidably occurs. However, an
additional factor, peculiar to the gate experiment, takes place -- a change of
V$_{g}$ brings into the system particles that are \textquotedblleft
memory-less\textquotedblright\ (as they are injected through \textit{metallic}
contacts where they are well equilibrated) thus \textquotedblleft
diluting\textquotedblright\ its memory. (This is somewhat analogous to the
effect of flushing a \textquotedblleft contaminated\textquotedblright\ vessel
with \textquotedblleft fresh\textquotedblright\ substance). All other things
being equal, this detrimental effect will be more pronounced in a low-density
system. Note that the value of $\Gamma$ for the sample shown in figure 9 is
larger by a factor of two than the sample studied by Vaknin et al. Since
$\Gamma$ increases monotonically with the carrier concentration \cite{6}, this
means that the sample in reference 10 has smaller n, and is therefore more
susceptible to this additional memory loss by a given $\delta$n. The relevance
of this picture can be tested by a Monte-Carlo simulation: An aging
`experiment' performed by changing just the random sites energies should yield
different results than by inserting new particles.

The degree of deviations from simple aging behavior, and the relevance of the
sample parameters could be likened to other phenomena where, due to a large
drive, the response gets out of the linear regime. Ohm's law, for example, is
obeyed when the energy imparted by the applied voltage to the charge carriers
is smaller than k$_{B}$T. The conductance will become voltage dependent when
the voltage is higher than this limit, but the degree of deviation from Ohm's
law will depend on many factors in particular, the temperature coefficient of
resistivity. Likewise, the linear-response regime (defined experimentally by
the condition that simple aging is strictly obeyed), of the aging experiment
has the limitation that $\delta$V$_{g}$/$\Gamma$ should not exceed a certain
value of order unity. We showed that the degree of deviation when this
condition is violated depends on the system.

The asymmetry created by a large $\delta$V$_{g}$ impairs the precise scaling
of the aging function $\Delta$G(t/t$_{w}$) by a smaller degree than that
revealed in the `learning-memory' experiments (compare figure 6 and 9). The
main reason is probably the cumulative effect produced in the latter type of
experiment. As mentioned before, the reason for the destructive effect is the
energy associated with $\delta$V$_{g}$ and therefore the `critical' $\delta
$V$_{g}$ at which the symmetry begins to fade away marks the typical energy of
the correlations. Inasmuch as the question of this important factor is
concerned, the `learning-memory' experiment is a much more sensitive tool than
the aging experiment. It would be of interest to apply a similar technique to
other glassy systems.

In summary, we presented and discussed experimental data pertaining to the
effect of history on relaxation processes in the electronic glassy phase of
indium-oxide samples. It was shown that the natural relaxation law in these
systems is logarithmic over a wide range of times provided the excitation is
performed following ample equilibration period. Partial equilibration, for a
finite t$_{w}$, leads to a deviation from the logarithmic law, beginning from
times t$\approx$t$_{w}$. This aging phenomenon takes a simple form such that
the relaxation of the excess conductance can be described by $\Delta
$G(t/t$_{w}$). It was further demonstrated that this simple-aging behavior
could still be observed (albeit somewhat impaired) even when $\delta$V$_{g}$
was large enough such that the symmetry in the dynamics of the `old' and `new'
states was considerably impaired. The robustness of the simple aging under
these seemingly unfavorable conditions is remarkable and deserves further studies.

This research was supported by a grant administered by the US Israel
Binational Science Foundation and by the German-Israeli Science Foundation.

\subsection{Figure Captions.}

1. The dependence of the conductance G on time following a quench from T=120K
to T$_{m}$=4.11K.

2. The conductance G as function of time after the gate voltage was changed
from 50V to +50V. Prior to this change, the sample was under V$_{g}$= 50V for
six days. The sample has R$_{\square}$=52M$\Omega$.

3. Same as in figure 2 except for the following history: The sample was
equilibrated under V$_{g}$=+50V for six days, then V$_{g}$ was switched to
-50V and maintained there for 1600 seconds before the final switch back to
+50V was affected. Note that the deviation of G(t) from the initial
logarithmic dependence (dashed line) is already evident after $\approx$300sec.

4. The dependence of the conductance on time while the gate voltage changes
periodically between two values differing by $\delta$V$_{g}$. The first change
occurs at the end of the short horizontal line at small t. The solid line is
for $\delta$V$_{g}$=100V, the dashed line for $\delta$V$_{g}$=400V, both on
the same sample (R$_{\square}$=52M$\Omega$).

5. Same as the dashed line ($\delta$V$_{g}$=400V) in figure 4 but on a log
time scale. The long-dashed line represents G(t) following the first change of
V$_{g}$. The short-dashed line (\textquotedblleft memory\textquotedblright%
\ curve) connects the dots just before the changes from \textquotedblleft
old\textquotedblright\ to \textquotedblleft new\textquotedblright\ V$_{g}$.
The thick solid line (\textquotedblleft learning\textquotedblright\ curve)
connects the dots just before the changes from \textquotedblleft
new\textquotedblright\ to \textquotedblleft old\textquotedblright\ V$_{g}$.
The difference in the (absolute) values of the slopes of these two lines is
the measure of asymmetry (see text).

6. The (absolute) value of the \textquotedblleft learning\textquotedblright%
\ and \textquotedblleft memory\textquotedblright\ slopes (in percent change of
$\Delta$G per decade in time) as function of $\delta$V$_{g}$. Circles and
triangles are for \textquotedblleft learning\textquotedblright\ and
\textquotedblleft memory\textquotedblright\ respectively. Full symbols data
points were taken by going from -Vg to +Vg and vice versa for the empty
symbols. The sample has R$_{\square}$=40M$\Omega$.

7. The ratio $\gamma$=q$_{F}$/q$_{L}$ for different values of $\delta$V$_{g}$
as a function of the transition rate between V$_{g}^{n}$ and V$_{g}^{o}$. The
dashed line in this figure indicates a particular choice of equal
transition-times. Sample with R$_{\square}$=52M$\Omega$.

8. An example of the cusp in G(V$_{g}$) (for the sample with R$_{\square}%
$=40M$\Omega$) measured using different sweep-rates. Note that the
characteristic width of the cusp $\Gamma$ ($\approx$50V) is independent of the
sweep-rate used (c.f., reference 6).

9. Two sets of \textquotedblleft aging\textquotedblright\ experiment each
using five identical values of waiting-times t$_{w}$. and employing two
different values of V$_{g}^{o}$ and V$_{g}^{n}$: (a) V$_{g}^{o}$ = -50V,
V$_{g}^{n}$=+50V (b) V$_{g}^{o}$ = -200V, V$_{g}^{n}$=+200V . Notice the
extrapolation of the logarithmic part to t$_{w}$ for the smaller $\delta
$V$_{g}$ but a failure to do so for the larger $\delta$V$_{g}$. Also, the
collapse of the data is somewhat worse in (b). Sample with R$_{\square}%
$=52M$\Omega$.


\begin{thebibliography}{99}                                                                                               %


\bibitem {1}M. Ben Chorin, Z. Ovadyahu and M. Pollak, Phys. Rev. \textbf{B48},
15025 (1993).

\bibitem {2}G. Martinez-Arizala, D. E. Grupp, C. Christiansen, A. Mack, N.
Markovic, Y. Seguchi, and A. M. Goldman, Phys. Rev. Letters, \textbf{78}, 1130
(1997). G. Martinez-Arizala, C. Christiansen, D. E. Grupp, N. Markovic, A.
Mack, and A. M. Goldman, Phys. Rev. \textbf{B57}, R670 (1998).

\bibitem {3}Z. Ovadyahu and M. Pollak, Phys. Rev. Letters, \textbf{79}, 459 (1997).

\bibitem {4}S. Bogdanovich and D. Popovic, Phys. Rev. Letters, \textbf{88},
23641 (2002); T. Grenet, Eur. Phys. J, \textbf{32}, 275 (2003).

\bibitem {5}J. H. Davies, P. A. Lee, and T. M. Rice, Phys. Rev. Letters, 49,
758 (1982); M. Pollak, Phil. Mag. \textbf{B50}, 265 (1984); M. Pollak and M.
Ortu\~{n}o, Sol. Energy Mater., \textbf{8}, 81 (1982); M. Gr\"{u}newald, B.
Pohlman, L. Schweitzer, and D. W\"{u}rtz, J. Phys. C, \textbf{15}, L1153 (1982).

\bibitem {6}A. Vaknin, Z. Ovadyhau, and M. Pollak, Phys. Rev. Letters,
\textbf{81}, 669 (1998).

\bibitem {7}The relevance of interactions to long relaxation times was
considered theoretically by: C. C. Yu, Phys. Rev. Lett., \textbf{82}, 4074 (1999).

\bibitem {8}L.C.E. Struik, Physical aging in amorphous polymers and other
materials (Elsevier, Amsterdam, 1978); L. M. Hodge, Science, \textbf{267},
1945 (1995).

\bibitem {9}A. Vaknin, Z. Ovadyhau, and M. Pollak, Phys. Rev. Letters,
\textbf{84}, 3402 (2000).

\bibitem {10}A. Vaknin, Z. Ovadyhau, and M. Pollak, Phys. Rev. \textbf{B63},
235403 (2002).

\bibitem {11}Z. Ovadyahu, J. Phys. C: Solid State Phys., \textbf{19}, 5187 (1986).

\bibitem {12}A. Vaknin, Z. Ovadyhau, and M. Pollak, Phys. Rev. \textbf{B61},
6692 (2000).

\bibitem {13}A. Vaknin, M. Pollak and Z. Ovadyahu, Springer Proc. in Physics
\textbf{87}, 995 (2001).

\bibitem {14}M. Pollak, Discuss. Faraday Soc. \textbf{50}, 11 (1970).

\bibitem {15}V. Orlyanchik, A Vaknin, and Z. Ovadyahu, Phys. Stat. Sol.,
\textbf{b239}, 67 (2002).
\end{thebibliography}
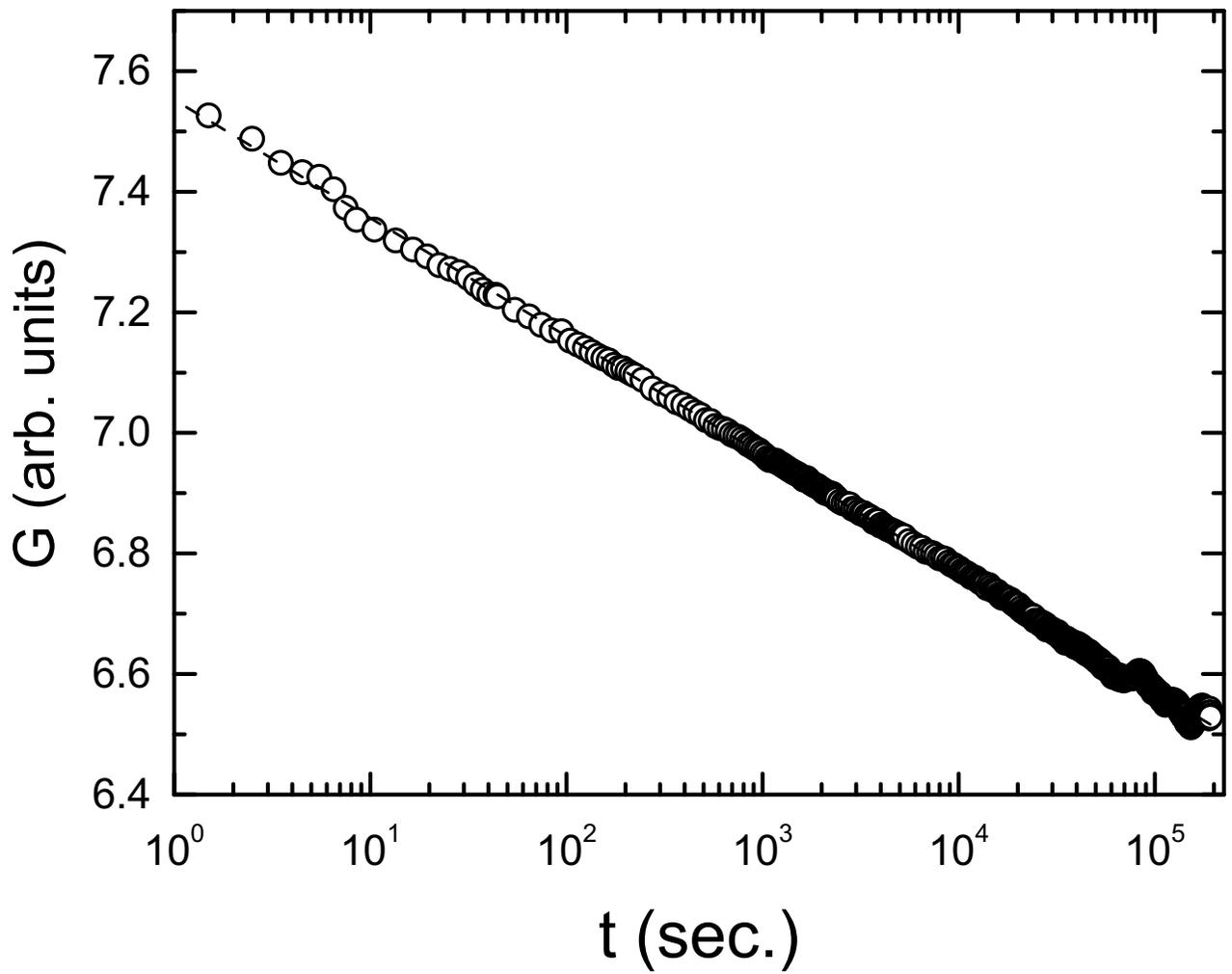
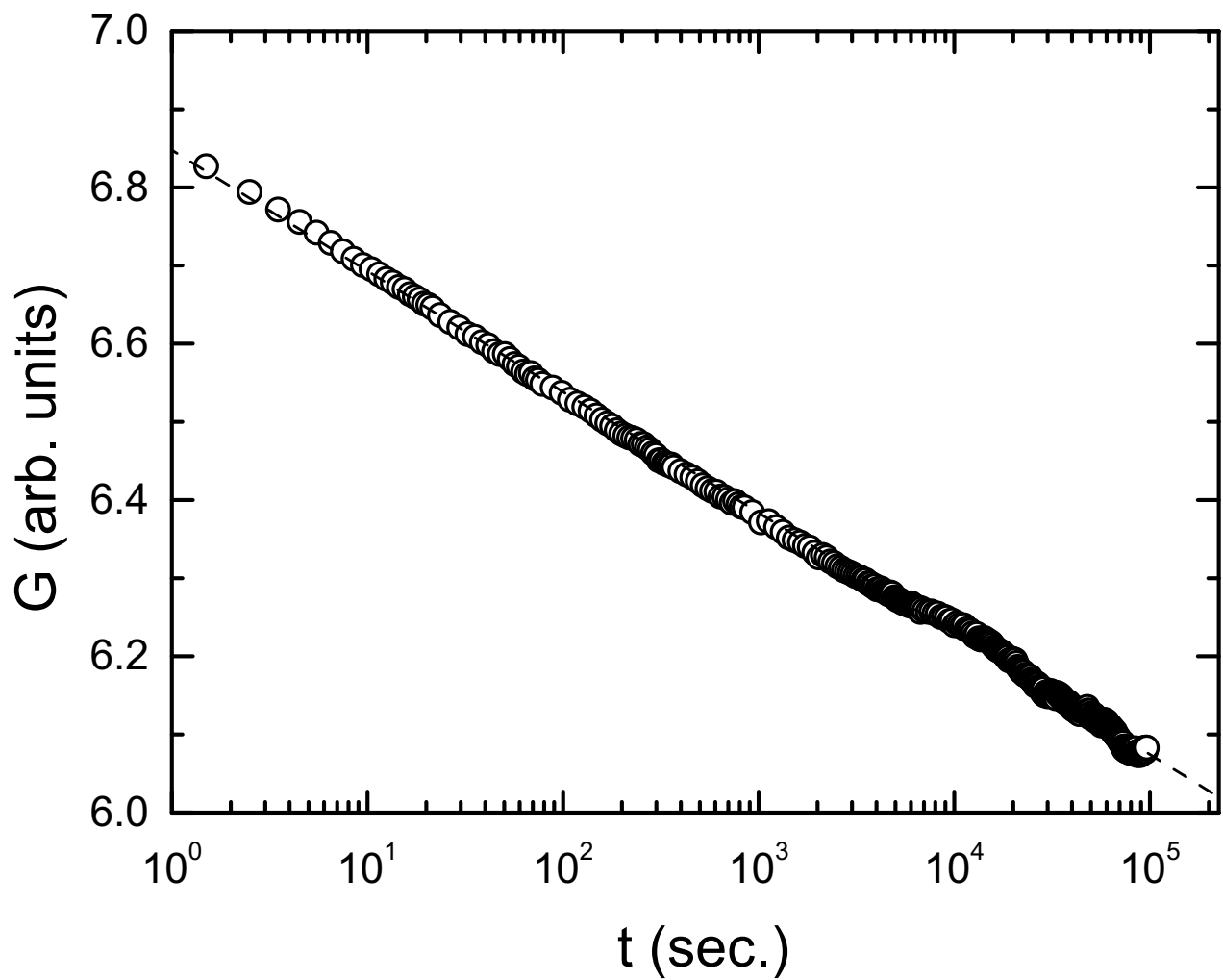
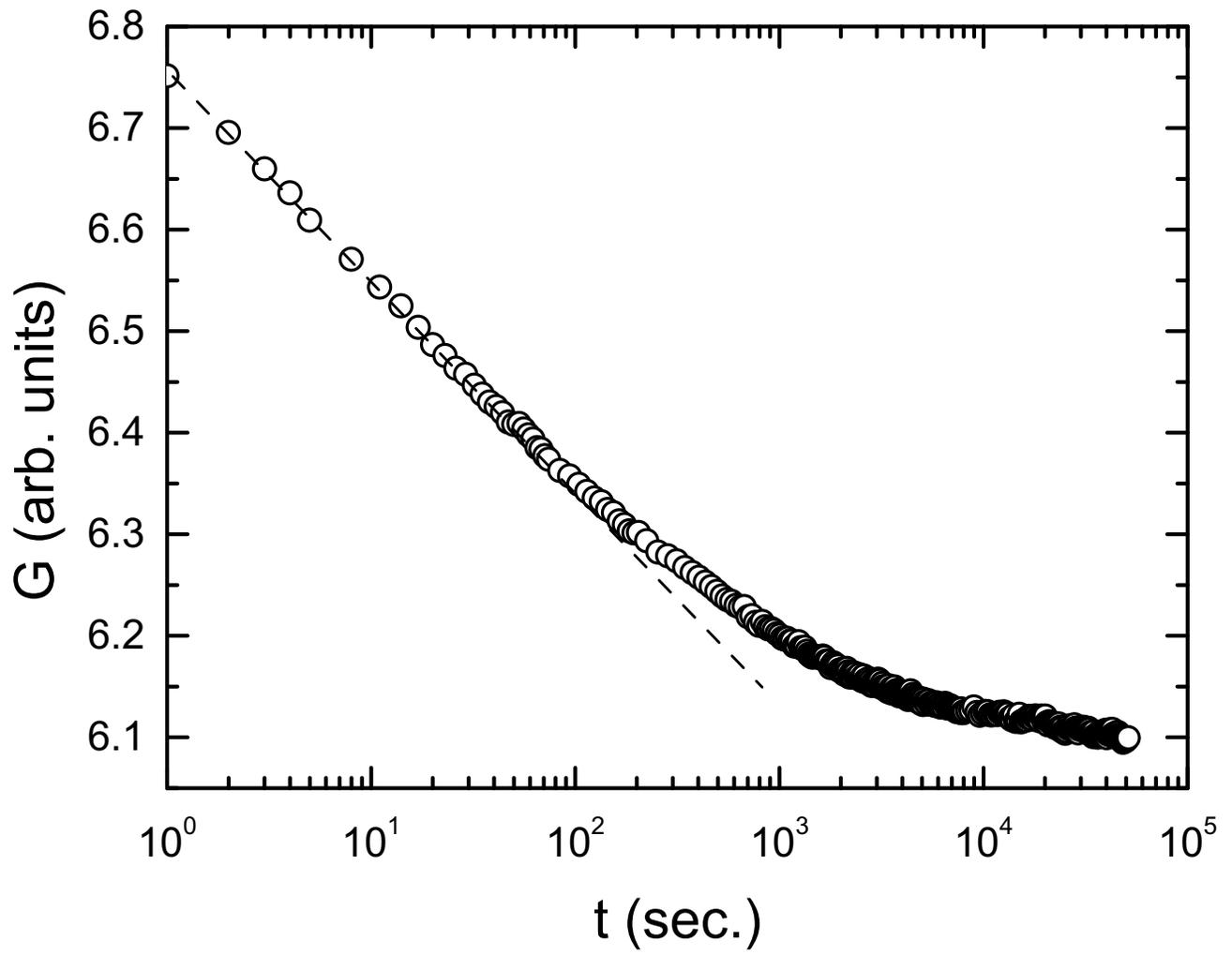
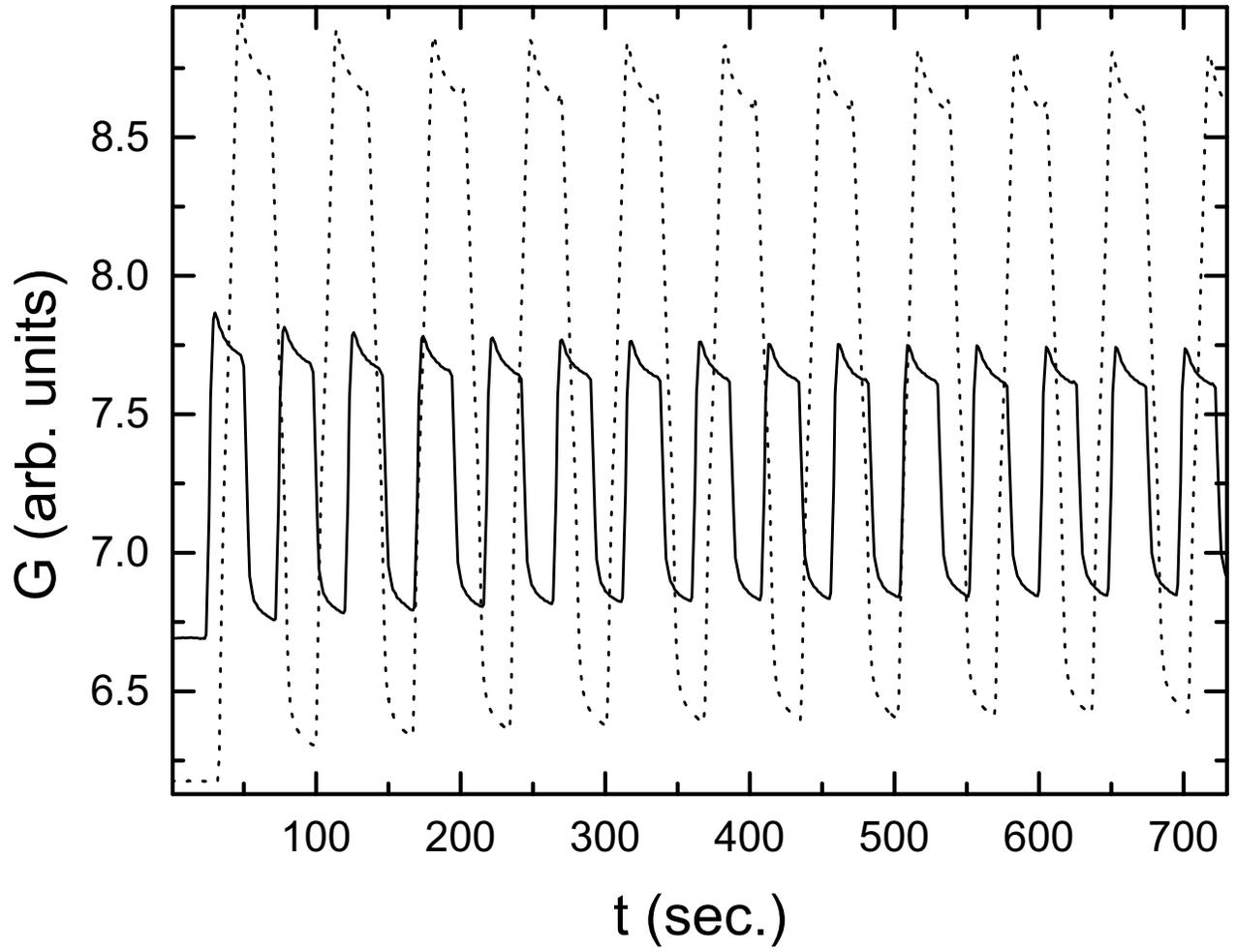
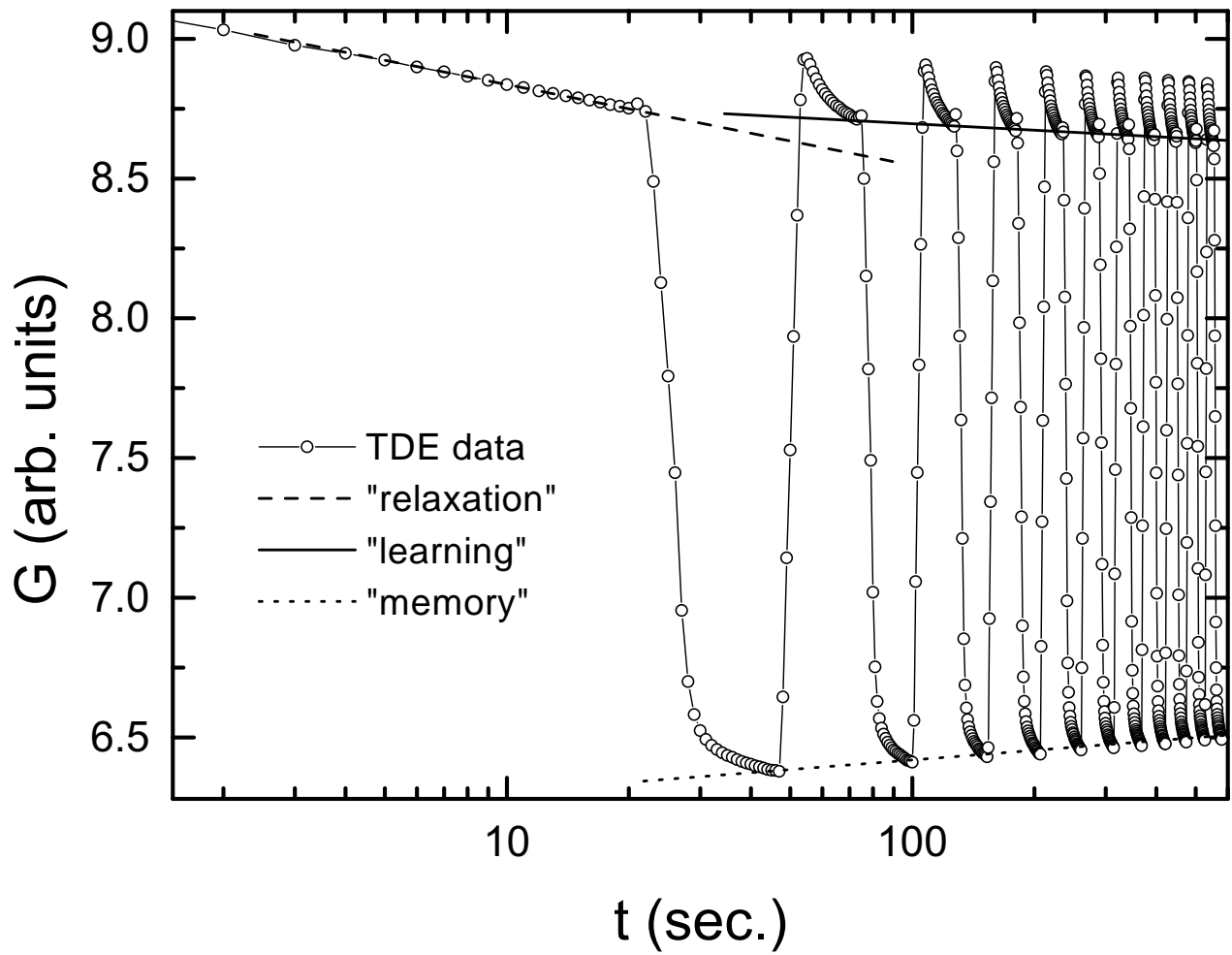
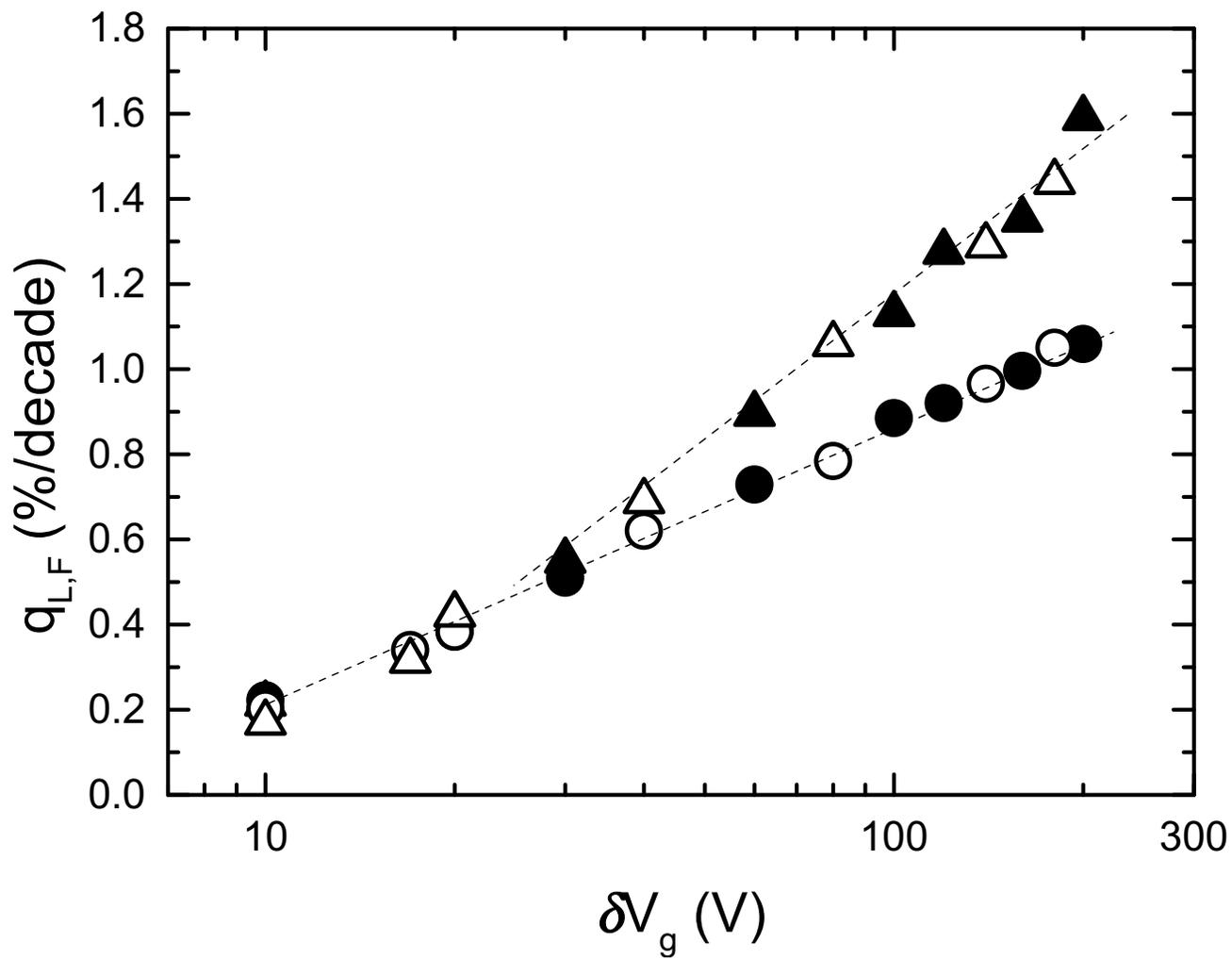
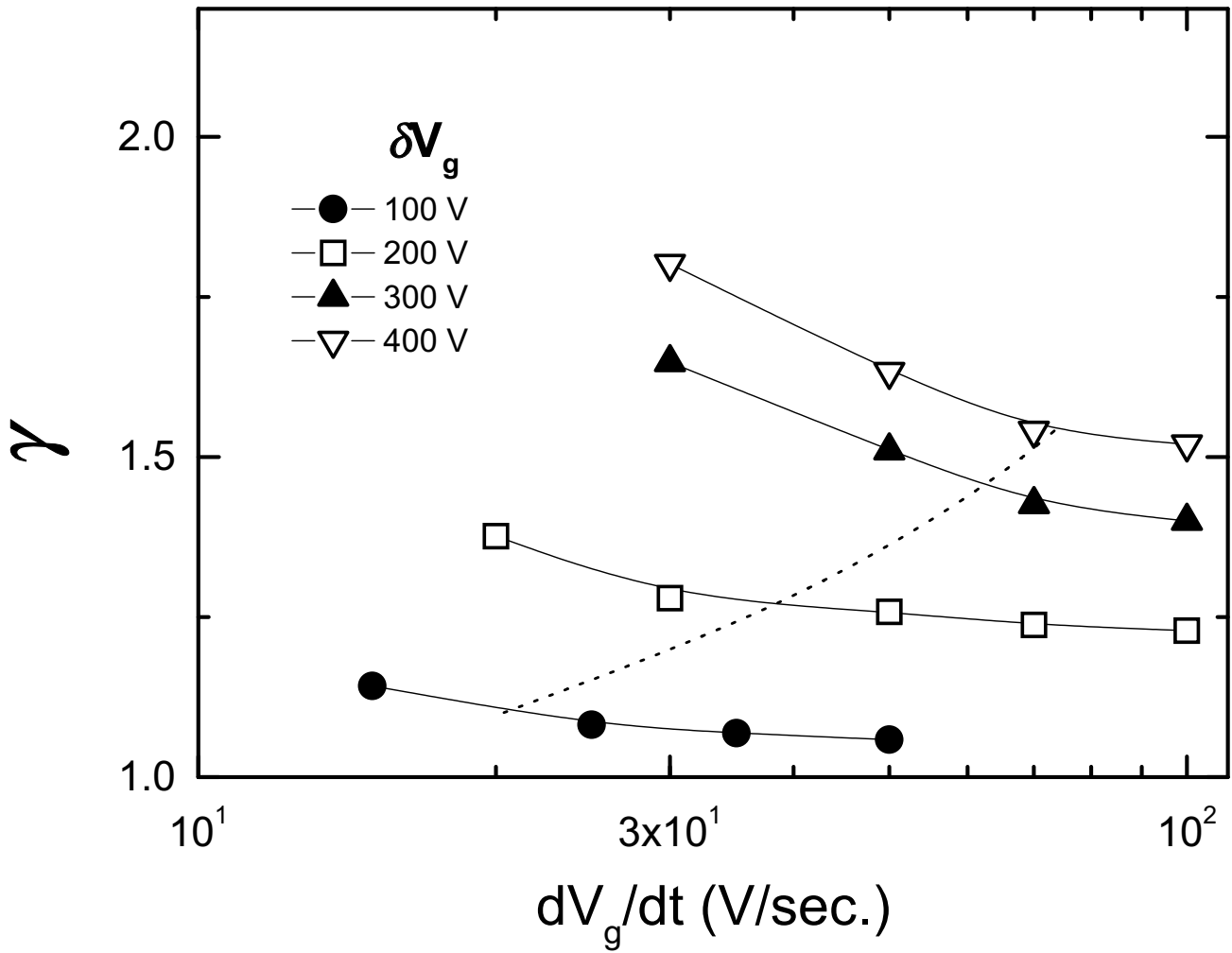
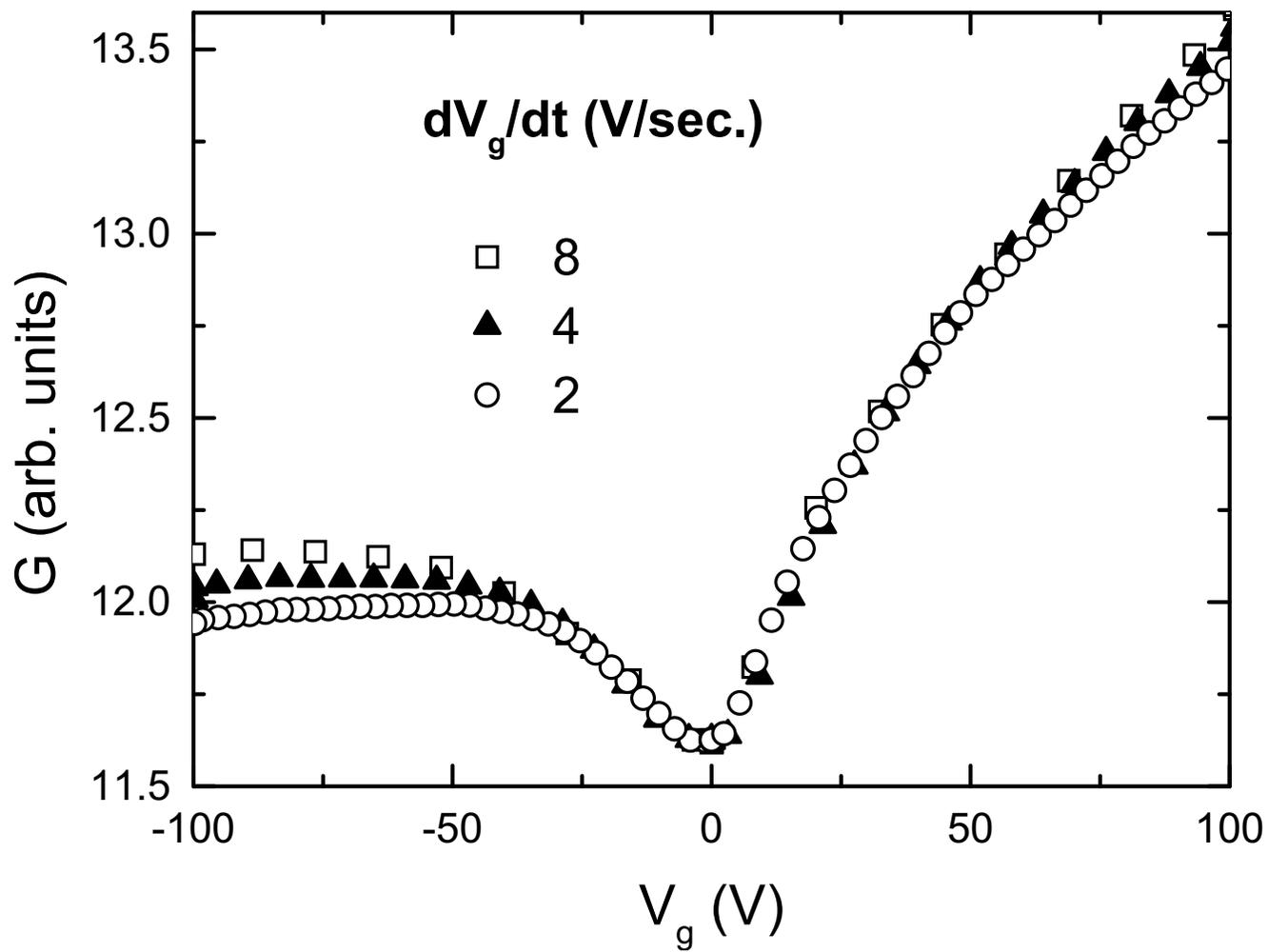
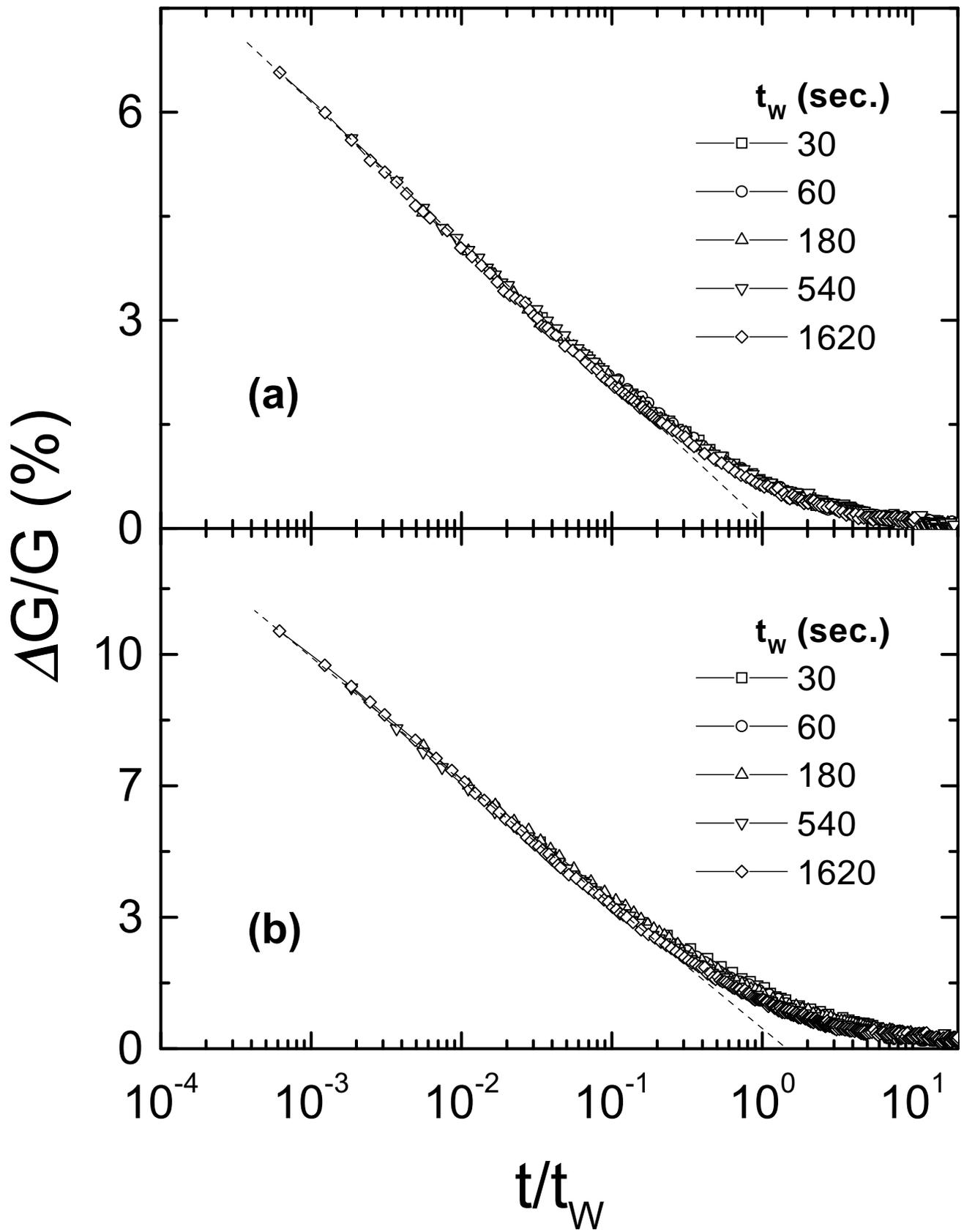
\end{document}